\documentclass[aps,prd,showpacs,twocolumn,
superscriptaddress]{revtex4-1}

\usepackage{graphicx}
\usepackage{subfigure} 
\usepackage{dcolumn}

\graphicspath{{figures/}}
\usepackage{amsmath}
\usepackage{amssymb}
\usepackage{bm}
\usepackage{color}
\usepackage{comment}

\usepackage[normalem]{ulem}
\usepackage[dvipsnames]{xcolor}
\usepackage{hyperref}
\hypersetup{
colorlinks=true, 
linkcolor=blue,  
citecolor=cyan,  
}

\newcommand{\bea}{\begin{eqnarray}}
\newcommand{\eea}{\end{eqnarray}}
\newcommand{\be}{\begin{equation}}
\newcommand{\ee}{\end{equation}}

\begin{document}
\title{Constraining the deformation of a black hole mimicker from the shadow
}
\author{Song Li}
\email{leesong@shao.ac.cn}
\affiliation{Shanghai Astronomical Observatory, Shanghai, 200030, China }
\affiliation{School of Astronomy and Space Science, University of Chinese Academy of Sciences,
Beijing, 100049, China}
\author{Temurbek Mirzaev}
\email{mtemur141096@gmail.com}
\affiliation{Ulugh Beg Astronomical Institute, Astronomy St. 33,
	Tashkent 100052, Uzbekistan }
\author{Ahmadjon A. Abdujabbarov}
\email{ahmadjon@astrin.uz}
\affiliation{Shanghai Astronomical Observatory, Shanghai, 200030, China }
\affiliation{Ulugh Beg Astronomical Institute, Astronomy St. 33,
	Tashkent 100052, Uzbekistan }
\author{Daniele~Malafarina}
\email{daniele.malafarina@nu.edu.kz}
\affiliation{Department of Physics, Nazarbayev University, 53 Kabanbay Batyr avenue, 010000 Astana, Kazakhstan }
\author{Bobomurat Ahmedov}
\email{ahmedov@astrin.uz}
\affiliation{Ulugh Beg Astronomical Institute, Astronomy St. 33,
	Tashkent 100052, Uzbekistan }
\affiliation{Tashkent Institute of Irrigation and Agricultural Mechanization Engineers, Kori Niyoziy, 39, Tashkent 100000, Uzbekistan}
\affiliation{Physics Faculty, National University of Uzbekistan, Tashkent 100174, Uzbekistan}
\author{Wen-Biao Han}
\email{wbhan@shao.ac.cn}
\affiliation{School of Fundamental Physics and Mathematical Sciences, Hangzhou Institute for Advanced Study, UCAS, Hangzhou 310024, China }
\affiliation{Shanghai Astronomical Observatory, Shanghai, 200030, China }
\affiliation{School of Astronomy and Space Science, University of Chinese Academy of Sciences,
Beijing, 100049, China}
\affiliation{International Centre for Theoretical Physics Asia-Pacific, Beijing/Hangzhou, China}
\affiliation{Shanghai Frontiers Science Center for  Gravitational Wave Detection, 800 Dongchuan Road, Shanghai 200240, China}
\date{\today}

\begin{abstract}
We consider a black hole mimicker given by an exact solution of the stationary and axially symmetric field equations in vacuum known as the $\delta$-Kerr metric. We study its optical properties 
based on a ray-tracing code for photon motion and characterize 
the apparent shape of the shadow of the compact object and compare it with the Kerr black hole.  
For the purpose of obtaining qualitative estimates related to the observed shadow of the supermassive compact object in the galaxy M87 we focus on values of the object's spin $a$ and inclination angle of observation $\theta_0$ close to the measured values.
We then apply the model to the shadow of the $\delta$-Kerr metric to obtain constraints on the allowed values of the deformation parameter. 
We show that based uniquely on one set of observations of the shadow's boundary it is not possible to exclude the $\delta$-Kerr solution as a viable source for the geometry in the exterior of the compact object.
\end{abstract}

\maketitle

\section{Introduction}

The recent observations of shadow images of the gravitationally collapsed objects at the center of the elliptical galaxy M87 (denoted by M87*) and at the center of the Milky Way galaxy (known as Sagittarius-A*, or in brief SgrA*) by the Event Horizon Telescope (EHT) collaboration \cite{EHT_Fi,EHT-SGR} has opened a new trend of research aimed to develop tests of gravity theories and corresponding black hole solutions near a regime of the gravitational field for which the validity General Relativity (GR) has never been tested \cite{tests1,tests2, tests3}. 
Consequently, for the first time we are on the verge of being able to probe the nature of the geometry in the vicinity of an astrophysical black hole candidate and test the validity of the Kerr hypothesis\cite{kerr-hypothesis1, kerr-hypothesis2,kerr-hypothesis3}.

What the EHT collaboration has observed is the relativistic image of the accretion disk surrounding the object, appearing due to light bending in the gravitational field of the source. The phenomenon has been thoroughly studied in GR and theoretical explanations of the relativistic images can be found for example in Refs.~\cite{Virbhadra2000, Virbhadra2009, BH_DC, Abdujabbarov_15}. 
The black hole shadow is related to the photon sphere around the central object and the conceptual study of the photon spheres is described for example in Ref.~\cite{Claudel2001}, 
while theoretical studies of the shadow of rotating black holes have been carried out in Refs.~\cite{Takahashi05, Bambi09, Hioki09, Amarilla10, Bambi10, Amarilla12, Amarilla13, Abdujabbarov13c, Wei13, Atamurotov13b, Ghasemi-Nodehi15, Cunha15, Quevedo11, Javed19, Ovgun19, Ovgun19a, Johannsen11, Younsi16, Ayzenberg18}. Also, observable quantities related to the black hole shadow have been investigated in detail in Refs.~\cite{Atamurotov15a, Ohgami15, Grenzebach15, Mureika17, Abdujabbarov17b, Abdujabbarov16a, Abdujabbarov16b, Mizuno18, Shaikh18b, Perlick17,Schee15, Schee09a, Stuchlik14, Schee09, Stuchlik10,Mishra19,Eiroa18, Giddings19, Ayzenberg18}.

Current observational data such as the observation of the shadow by EHT and the waveforms of detected gravitational waves by the LIGO-Virgo collaborations \cite{LIGO_a, LIGO_b, LIGO_c, LIGO_d}, cannot yet provide certainty that the observed astrophysical black hole candidates are exactly described by the Kerr metric. 
When considering other possibilities one may encounter a degeneracy when the same values for observational quantities may be obtained from different and distinct scenarios
which may therefore mimic the Kerr space-time~\cite{Carballo-Rubio18,Abdikamalov19}. 
Nevertheless one may devise other methods to obtain data from the proposed solutions for compact objects within GR and/or modified theories of gravity to constrain the physical validity of such models and possibly restrict or break the aforementioned degeneracies~\cite{Bambi17c,Berti15,Cardoso16a,Cardoso17,Krawczynski12, Krawczynski18, Yagi16, Bambi16b, Yunes13, Tripathi19}.  

The data from EHT observations can be used to get constraints on the parameters of the mathematical solutions that could describe the geometry surrounding the objects M87* and SgrA*. These solutions range from black hole space-times in modified and alternative theories of gravity \cite{Nodehi2020, Jusufi2021, Liu2020, Jusufi2020, Afrin2021}, to black hole mimickers and exotic compact object in GR \cite{boson1, boson2, boson3, Abdikamalov19, joshi1, joshi2} and classical GR black hole with hair or immersed in matter fields \cite{hair1, hair2, hair3, hair4, hair5, dm1, dm2, dm3}. In fact it is indeed possible, as we shall show in this paper, that the geometry is described by an exact solution of the vacuum Einstein's equations which is not a black hole.
 
In the present article we study the shadow of a rotating black hole mimicker described by the line element proposed in~\cite{quevedo} and subsequently studied in~\cite{deltaKerr}, known as $\delta$-Kerr metric. This is an exact solution of the vacuum field equations that extends the Kerr solution to include higher mass multipole moments.
The Kerr space-time, which describes the gravitational field of an axially-symmetric rotating black hole, is fully characterized by two parameters, i.e. the mass $m$ and the specific angular momentum $a$. The $\delta$-Kerr space-time is a generalization of the Kerr space-time which belongs to the class of Ricci-flat vacuum solutions of general relativity discussed in Ref.~\cite{quevedo} and is obtained from the Zipoy-Voorhees static solution~\cite{Zipoy, Zipoy_66, Voorhees} by means of a solution generating technique known as HKX transformation \cite{HKX}, due to Hoenselaers, Kinnersley and Xanthopoulos.

The Zipoy-Voorhees space-time, sometimes called $\delta$-metric or $\gamma$-metric or $q$-metric, has been extensively studied over the past several years \cite{ZV1,ZV2,ZV3,ZV4,ZV5,ZV6,ZV7,ZV8}, however it is not ideal to describe an astrophysical source due to its static nature. The nonlinear superposition of the Zipoy-Voorhees metric and the Kerr metric results in $\delta$-Kerr metric that represents a `deformed' Kerr solution which is characterized by three parameters, namely the mass $m$, the spin $a$ and the deformation parameter $\delta$, and it generically possesses naked singularities.
The $\delta$-Kerr space-time and its relativistic multipole moments have been recently explored in Ref.~\cite{Toktarbay14} and \cite{Frutos_Alfaro18}. 
The parameter $\delta$ describes the departure from the black hole solution with $q=1-\delta $ related to the non relativistic mass quadrupole moment of the source. 

In this article we study the image of the shadow of the $\delta$-Kerr metric and investigate whether the results of the EHT collaboration for M87* may be used to constrain the value of the deformation parameter. We find that for oblate sources (i.e. $q>0$), for any given value of the deformation parameter close to Kerr, there is always a value of the object's spin that produces the same boundary of the shadow within the experimental bounds obtained by the EHT collaboration. Therefore, it may not be possible to distinguish the $\delta$-Kerr space-time from its black hole counterpart if one relies uniquely on one kind of observations, such as the shadow's image.

The paper is organized in the following way: In Section~\ref{metric} the properties of $\delta$-Kerr space-time are briefly described. 
Section~\ref{rtcs} is devoted to description of the ray-tracing code for photons while the apparent shape of the shadow of the compact object and its relation with the observed border of the shadow of M87* are discussed in Section~\ref{evaluation}. 
In Section~\ref{disk} we obtain the image of the thin accretion disk in the $\delta$-Kerr space-time and
finally, Section~\ref{Summary} summarizes the main results obtained and the conclusions that can be drawn in view of future observations. 

\section{The $\delta$-Kerr space-time}\label{metric}

Our aim is to consider a geometry that is continuously linked to the Kerr space-time and is also an exact solution of the vacuum field equations. Therefore the most natural choice is to consider a member of the class of stationary and vacuum exact solutions, known as Weyl class. 
The extension of black hole solutions by the inclusion of higher mass multipole moments takes into account possible deformations from the black hole geometry and, in the simplest case, it amounts to a single additional parameter related to the mass quadrupole. In our analysis we shall consider the so called $\delta$-Kerr space-time. Following \cite{deltaKerr} the $\delta$-Kerr metric in Boyer-Lindquist coordinates $\{t,r,\theta,\phi\}$ takes the form:
\begin{equation}
\begin{aligned}
d s^{2}=&-F d t^{2}+2 F \omega d t d \phi+\frac{e^{2 \gamma}}{F} \frac{\mathbb{B}}{\mathbb{A}} d r^{2}+r^{2} \frac{e^{2 \gamma}}{F} \mathbb{B} d \theta^{2} \\
&+\left(\frac{r^{2}}{F} \mathbb{A} \sin ^{2} \theta-F \omega^{2}\right) d \phi^{2}\ , 
\end{aligned}
\end{equation}
where
\begin{equation}
\mathbb{A}=1-\frac{2 m}{r}+\frac{a_*^{2}}{r^{2}}\ , \quad \mathbb{B}=\mathbb{A}+\frac{\sigma^{2} \sin ^{2} \theta}{r^{2}}\ ,
\end{equation}
\begin{equation}
\begin{aligned}
&F=\frac{\mathcal{A}}{\mathcal{B}}, \quad \omega=2\left(a_*-\sigma \frac{\mathcal{C}}{\mathcal{A}}\right)\ , \\
&e^{2 \gamma}=\frac{1}{4}\left(1+\frac{m}{\sigma}\right)^{2} \frac{\mathcal{A}}{\left(x^{2}-1\right)^{\delta}}\left(\frac{x^{2}-1}{x^{2}-y^{2}}\right)^{\delta^{2}}\ .
\end{aligned}
\end{equation}
With
\begin{equation}
\begin{aligned}
&\mathcal{A}=a_{+} a_{-}+b_{+} b_{-}\ , \\
&\mathcal{B}=a_{+}^{2}+b_{+}^{2}\ , \\
&\mathcal{C}=(x+1)^{q}\left[x\left(1-y^{2}\right)(\lambda+\eta) a_{+}+y\left(x^{2}-1\right)(1-\lambda \eta) b_{+}\right]\ , 
\end{aligned}
\end{equation}
and we have defined
\bea
a_{\pm}&=&(x \pm 1)^{q}[x(1-\lambda \eta) \pm(1+\lambda \eta)]\ , \\
b_{\pm}&=&(x \pm 1)^{q}[y(\lambda+\eta) \mp(\lambda-\eta)] \ , \\
\lambda&=&\alpha(x^2-1)^{-q}(x+y)^{2q} \ , \\
\eta&=&\alpha(x^2-1)^{-q}(x-y)^{2q} \ ,
\eea
and for simplicity used prolate spheroidal coordinates $\{x,y\}$ that relate to $\{r,\theta\}$ via 
\begin{equation}
x=\frac{r-m}{\sigma}, \quad y=\cos \theta\ .
\end{equation}

The $\delta$-Kerr space-time depends on three free parameters, that are related to mass, angular momentum and deformation. The mass parameter $m$ is related to the total gravitational mass of the source.
The angular momentum parameter $a_*$ is analogous to Kerr's angular momentum and subject to the condition $|a_*| \leq m$. In the following it will be convenient to use the a-dimensional parameter $a=a_*/m$ that has range $a\in(-1,1)$. The deformation parameter $\delta$ is analogous to the deformation parameter in the static Zipoy-Voorhees metric. 

The other constants used above, namely $\sigma>0$, $q$ and $\alpha$ can all be related to the three parameters as follows:
\bea
\sigma&=&\sqrt{m^2-a_*^2}=m\sqrt{1-a^{2}}\ , \\
q&=&\delta-1 \ , \\
\alpha&=&\frac{m-\sigma}{a_*}=\frac{a_*}{m+\sigma}=\frac{a}{1+\sqrt{1-a^{2}}}\ . 
\eea
So we can interpret $q$ as related to the quadrupole parameter and $\alpha$ as a new definition of the rotation parameter.

The $\delta$-Kerr is asymptotically flat and has the following limits:
\begin{itemize}
    \item[-] For $\delta=1$ one obtains the Kerr solution.
    \item[-] For $a=0$ one obtains the Zipoy-Voorhees metric.
    \item[-] For both $\delta=1$ and $a=0$ one obtains the Schwarzschild space-time.
\end{itemize}
Furthermore the mass monopole and mass quadrupole are given by
\bea \label{m0}
 M_0&=&  m+\sigma q=m(1+q\sqrt{1-a^2}) \ , \\ 
 M_2&=& \frac{\sigma^3 q}{3}(7-q^2)+m\sigma^2(1-q^2)-m^2(m+3\sigma q) \ ,
\eea 
which reduce to the known values for the Zipoy-Voorhees metric when $a=0$ and the known values for Kerr when $q=0$.
Also, the angular monopole and angular quadrupole are given by (see \cite{Toktarbay14})
\bea \label{j1}
 J_1&=& ma_*+2a_*\sigma q=m^2a(1+2q\sqrt{1-a^2}) \ ,  \\ 
 J_3&=& -\frac{a_*}{3}[3a_*^2m+3m\sigma q(3\sigma q+4m)+2\sigma^3 q (q^2-4)]\ .
\eea 
which vanish for $a=0$ and reduce to the known values for Kerr when $q=0$, i.e. when $\delta=1$.

The geometry of the $\delta$-Kerr space-time is particularly interesting because it is continuously linked to the widely studied black hole solutions and because, unlike black holes, it does not posses an event horizon. In fact it can be shown that the $\delta$-Kerr metric generically exhibits strong curvature singularities at various locations for $r\leq r_s$, with 
\be\label{singularity}
r_s= m+\sigma \ ,
\ee
corresponding to the location of the outer horizon in the Kerr case. For this reason, in order to consider the line element as describing the geometry outside a viable rotating compact object, we will restrict the range of the radial coordinate to $r>r_s$.

If we consider the possibility that the classical relativistic description breaks down at the horizon level and the final fate of complete gravitational collapse can be an exotic compact object which retains some of the higher order non Kerr multipole moments of its progenitor, then we might consider the $\delta$-Kerr solution as the exterior of such an exotic massive source.

In this regard the $\delta$-Kerr solution may be considered as a black hole mimicker and it important to study its observational properties in view of the possibility of testing the geometry in the vicinity of astrophysical black hole candidates in the future.  

\section{Ray-tracing and shadow's boundary}\label{rtcs}

Observational information of many astrophysical compact sources such as the supermassive black hole candidate in M87 comes from the study and analysis of the light emitted by the accretion disk surrounding the source and reaching the observer. Therefore, to study the observational properties of the sources of the $\delta$-Kerr space-time we need to model the light emission of the accretion disk. 
In the following, we study the trajectories of photons in the $\delta$-Kerr space-time applying the ray-tracing code described in Ref.~\cite{Ray_code}. 

The stationary and axisymmetric $\delta$-Kerr metric is independent of the coordinates $t$ and $\phi$, so there exist two Killing vectors, one timelike and one spacelike associated with time and rotation invariance about the symmetry axis $z$. Consequently, these two Killing vectors correspond to two conserved quantities, namely the energy $E$ and the z-component of the angular momentum $L_z=L$ of the particle, in our case a photon. The  two conserved quantities can be expressed as:
\begin{eqnarray}\label{conserved_E}
-E &=g_{t t} \dot{t}+g_{t \phi} \dot{\phi}\ ,\\
L &=g_{\phi t} \dot{t}+g_{\phi \phi} \dot{\phi}\ ,\label{conserved_L}
\end{eqnarray}
where the overhead dot represents the derivative with respect to the affine parameter of the trajectory $\lambda$.

To determine the evolution of the photon's trajectory we need the four equations of motion for $x^\mu(\lambda)$. Using the conserved quantities we obtain two first-order differential equations for the $t$ and $\phi$ components from equations~(\ref{conserved_E})-(\ref{conserved_L}). These can be rewritten in terms of the normalized affine parameter $\lambda'=E\lambda$ as  
\begin{equation}\label{dt}
\frac{d t}{d \lambda^{\prime}}=\frac{-g_{\phi \phi}-b g_{t \phi}}{g_{\phi \phi} g_{t t}-g_{t \phi}^{2}}\ ,
\end{equation}
\begin{equation}\label{dphi}
\frac{d \phi}{d \lambda^{\prime}}=\frac{b g_{t t}+b g_{t \phi}}{g_{\phi \phi} g_{t t}-g_{t \phi}^{2}}\ ,
\end{equation}
where $b$ is the photon's impact parameter which is defined as $b \equiv |L/E|$.

We can use the usual second-order geodesic equations with the normalized affine parameter $\lambda'$ and the Christoffel symbols $\Gamma^\sigma_{\mu\nu}$ to get remaining two equations for the $r$ and $\theta$ components of the photon's trajectory in the $\delta$-Kerr space-time. 
This way we obtain the full equations of motion that the ray-tracing code uses to determine the photon's trajectory.

\begin{figure}
	\includegraphics[width=0.5 \textwidth]{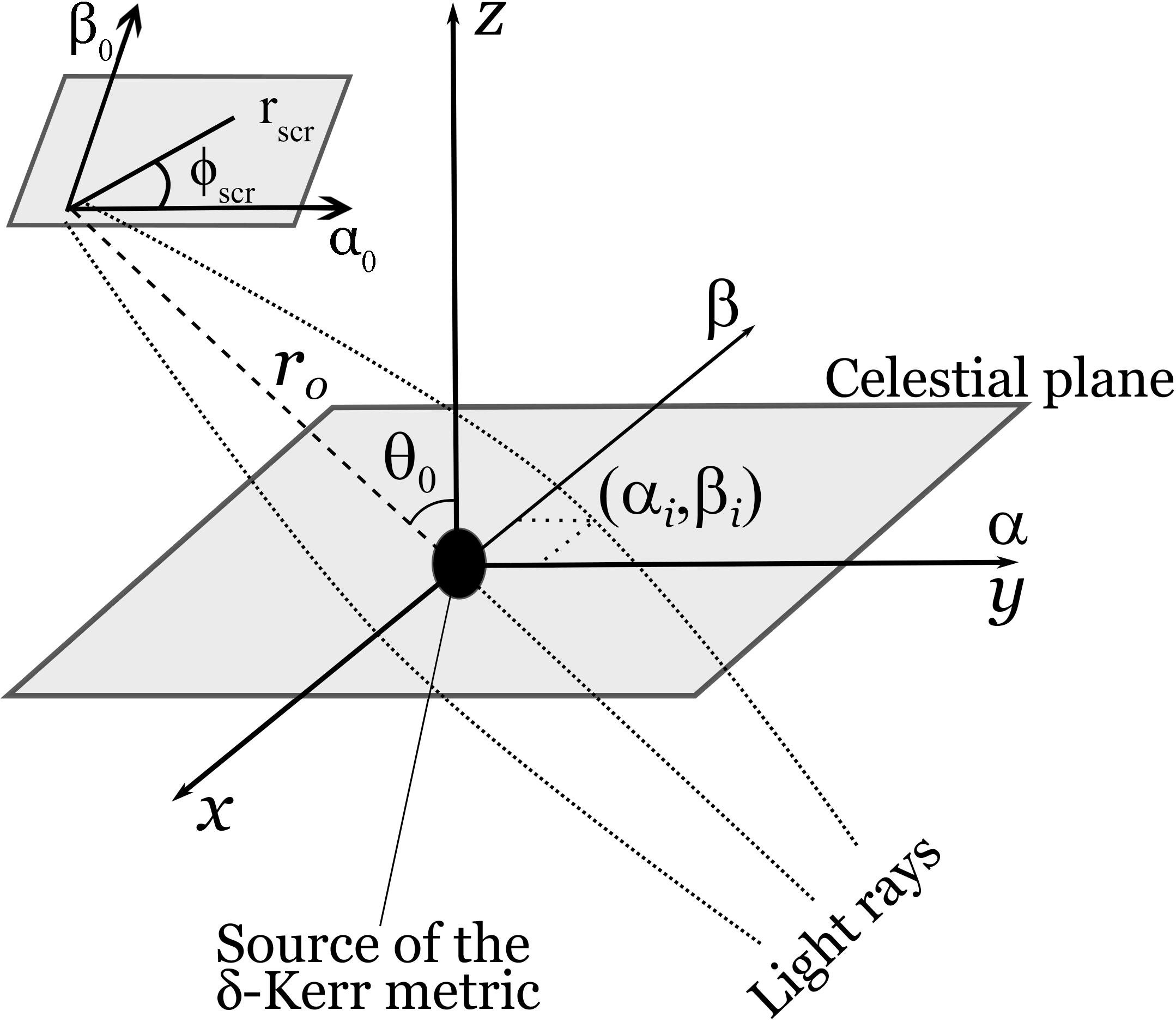}
	\caption{Schematic illustration of the celestial coordinates used for the ray tracing code in the $\delta$-Kerr space-time. \label{illustration}}
\end{figure}

We suppose that the source of the $\delta$-Kerr space-time is located at the center of the coordinate system. For convenience, we choose the mass parameter of the central object as $m=1$, since it just scales the size of the shadow and it does not affect its shape. 
We assume the screen to be at a large distance $r_0$ from the source (in the code we set $r_0=1000$), perpendicular to the line of sight with the location of the source and inclined by an angle $\theta_0$ with respect to the rotation axis (see Fig. \ref{illustration}).
The celestial coordinates $\{\alpha_0, \beta_0\}$ on the observer's sky are related to polar coordinates $r_{\rm scr}$ and $\phi_{\rm scr}$ on the screen by $\alpha_0=r_{\rm scr}\cos(\phi_{\rm scr})$ and $\beta_0=r_{\rm scr}\sin(\phi_{\rm scr})$. 
Because we only know the positions and momenta of the photon on the screen, we should solve the geodesic equations backwards from the observer's screen to the source. 
The photon departs from the screen with a four-momentum perpendicular to it and initial conditions $(\alpha_0,\beta_0)$. Using the equation $u^\alpha u_\alpha = 0$ of the photon's four-velocity one can find the component $(dt/d\lambda')_i$. The initial conditions of the photon on the screen are then converted to the initial position in the $\delta$-Kerr metric $(r_i,\theta_i,\phi_i)$.
The initial position and four-momentum of each photon in the $\delta$-Kerr space-time are given as~\cite{source_screen}
\begin{equation}\label{in_pos_r}
		 r_i =\left(r_0^2+\alpha_0^2+\beta_0^2\right)^{1/2}      ,\\
\end{equation}
\begin{equation}\label{in_pos_theta}
		 \theta_i=\arccos\left(\frac{r_0\cos\theta_0+\beta_0\sin\theta_0}{r_i}\right),\\
\end{equation}
\begin{equation}\label{in_pos_phi}
		 \phi_i = \arctan\left(\frac{\alpha_0}{r_0\sin\theta_0-\beta_0\cos\theta_0}\right),
\end{equation}
and
\begin{equation}\label{in_momen_r}
	 \left(\frac{dr}{d\lambda'}\right)_i=\frac{r_0}{r_i},\\
\end{equation}
\begin{equation}\label{in_momen_theta}
	 \left(\frac{d\theta}{d\lambda'}\right)_i=\frac{r_0(r_0\cos\theta_0+\beta_0\sin\theta_0)-r_i^2\cos\theta_0}{r_i^2\sqrt{r_i^2-(r_0\cos\theta_0+\beta_0\sin\theta_0)^2}},\\
\end{equation}
\begin{equation}\label{in_momen_phi}
	 \left(\frac{d\phi}{d\lambda'}\right)_i=\frac{-\alpha_0\sin\theta_0}{\alpha_0^2+(r_0\cos\theta_0+\beta_0\sin\theta_0)^2}.\\
\end{equation}

The initial conditions of the code on the screen are defined in the following way. The image of the compact source is confined inside $0 \leq r_{\rm scr} \leq 20$ and
$0 \leq \phi_{\rm scr} \leq 2\pi$ with a step of $\pi / 180$. 
On the screen we can define the boundary between the photons that can be captured by the compact source and the photons that can escape to infinity. The photons are captured by the compact source if they cross inside the surface $r = r_s + \Delta r$ with $\Delta r = 10^{-3}$ and $r_s$ being the radius of the infinitely redshifted null surface given by equation \eqref{singularity}. The confine is defined to an accuracy of $\Delta r_{\rm scr} = 10^{-3}$ which is enough to accurately determine the shadow's boundary with the value of $r_{\rm scr}$ and $\phi_{\rm scr}$. 
This method allows one to calculate the border of the shadow produced by light rays in the $\delta$-Kerr as it appears on the observer's screen within the accuracy defined.

The object's shadow is the observable consequence of the photon's capture by the massive compact source. In black hole space-times this is related to the photon sphere which is the surface formed from the combination of all unstable spherical photon orbits, namely the surface that separates between photon geodesics that are able to escape to spatial infinity and those that fall towards the compact object. 
To describe the shadow one may also consider the celestial coordinates $\alpha$ and $\beta$~(see \cite{Abdujabbarov16b} for reference) which are defined as,
\begin{eqnarray}
\alpha &=&\underset{r_0 \rightarrow \infty}{\lim}\left(-r_0^2 \sin \theta_0 \frac{d\phi}{dr}\right) \ , \label{eq:14}\\
\beta &=&\underset{r_0 \rightarrow \infty}{\lim}\left(r_0^2 \frac{d\theta}{dr}\right) \ , \label{eq:15}
\end{eqnarray}
where $r_0$ is the distance between the observer and massive source and $\theta_0$ is the inclination angle between the normal of observer's screen plane and the symmetry axis~(see Fig.~\ref{illustration}).

\begin{figure*}
\centering
\includegraphics[width=1.00 \textwidth]{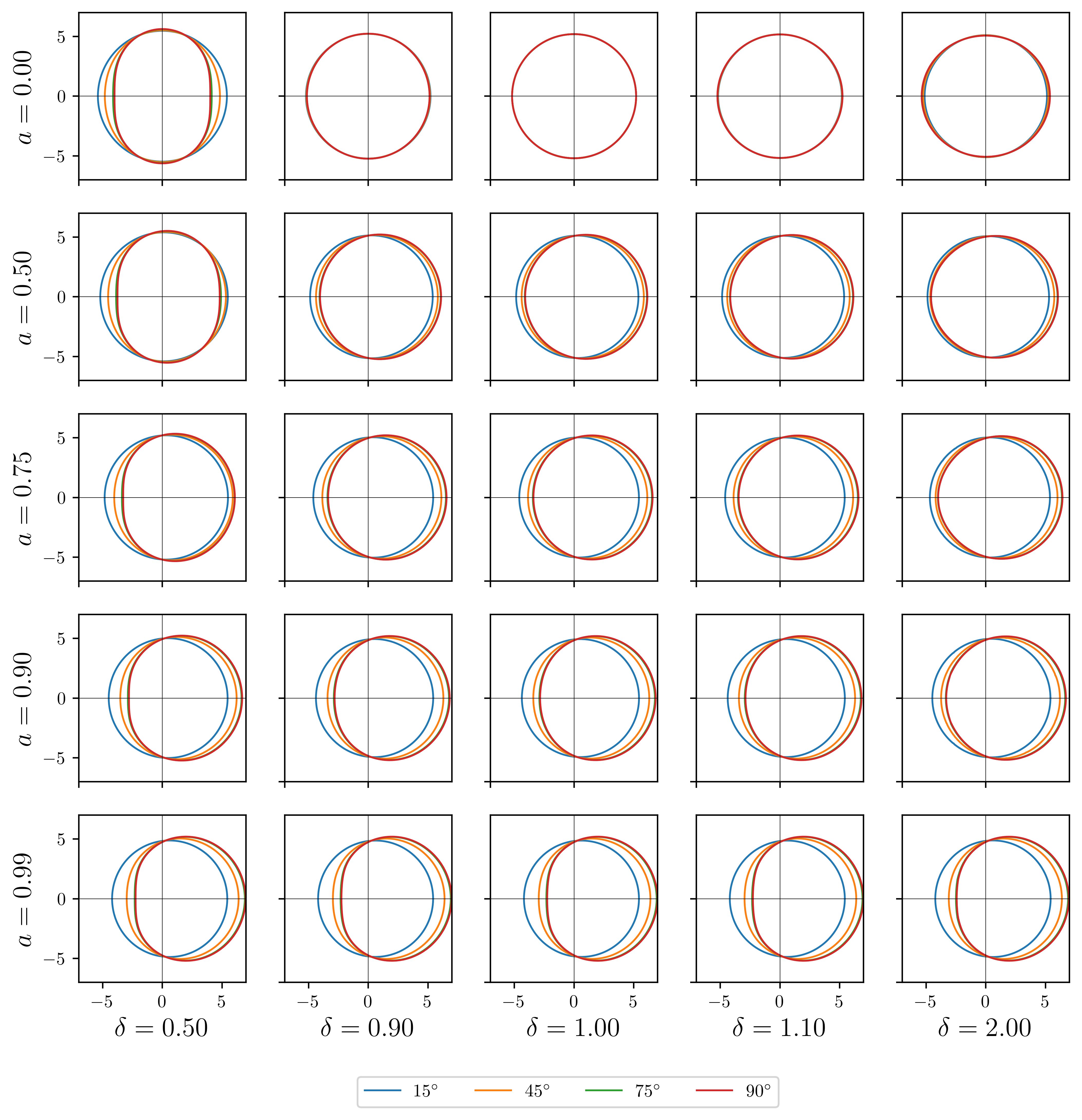}
\caption{The boundary of the shadow in the $\delta$-Kerr space-time with different values of spin $a$ and deformation $\delta$ for different inclination angle of observation $\theta_0$. Each panel shows the shadow's boundary for $\theta_0=15^{\circ},45^{\circ},75^{\circ},90^{\circ}$. For smaller values of $a$ the effect of $\delta$ on the shadow's boundary is more noticeable.}\label{Shadow_new1}.
\end{figure*}

Through this method we can get the boundary of the shadow as shown in Fig.~\ref{Shadow_new1}. As expected, the boundary of the shadow depends on the deformation parameter $q$, the spin parameter $a$ and the observer's inclination angle $\theta_0$. Notice that for larger values of $a$ the effects of the deformation parameter become less pronounced and are most visible for $q<0$ in the static case. This can be understood by looking at equations \eqref{m0} and \eqref{j1} which show how the leading mass and angular multipoles $M_0$ and $J_1$ depend on $q$ via $\sqrt{1-a^2}$.
Notice that for astrophysical rotating sources one would expect the rotation to induce some oblateness in the source thus causing a deformation with $q>0$. Fig.~\ref{Shadow_new1} allows us to estimate the influence of the mass quadrupole, via the deformation parameter $q$, on the shadow's border. In the sequel, we will focus on how we may constrain the value of the deformation parameter in the $\delta$-Kerr metric from observations.

\section{Constraining the value of the deformation parameter}\label{evaluation}

Can we estimate the deformation parameter of the $\delta$-Kerr space-time from observations? In particular, considering the image of the BH candidate in the galaxy M87, would it be possible to put constraints on the value of the deformation parameter?
In order to address this question we need to describe the dependence of the shape of the shadow on $q$ and relate it to the shadow's contour obtained experimentally. To this aim we will use the independent coordinate formalism proposed in~\cite{Ayzenberg18}. The shape of the shadow is parametrized in terms of the average radius of the sphere $\langle R \rangle$, and the asymmetry parameter $A$. We can safely ignore the shift $D_c$ off the center of the shadow because, in the case of the $\delta$-Kerr metric $D_c$ is always identically equal to zero. There are some other ways to describe the shape of the shadow (see e.g. \cite{Tsukamoto14, Abdujabbarov17b, Abdujabbarov_15}), however, the results are similar with any chosen parametrization. The average radius $\langle R \rangle$ is the average distance of the boundary of the shadow from its center, which is defined as
\begin{eqnarray}
\langle R \rangle \equiv \frac{1}{2\pi} \int_{0}^{2\pi} R(\vartheta)d\vartheta,
\end{eqnarray}
where $R(\vartheta) \equiv \left[(\alpha-D_c)^2+\beta(\alpha)^2\right]^{1/2}$, with $D_c=0$ and $ \vartheta \equiv \tan^{-1}[\beta(\alpha)/\alpha]$. The asymmetry parameter $A$ represents the distortion of the shadow from a circle and it is defined by
\begin{eqnarray}
A \equiv 2\left[\frac{1}{2\pi}\int_{0}^{2\pi} \left(R-\langle R \rangle \right)^2 d\vartheta \right]^{1/2}.
\end{eqnarray}

There are two methods that we can follow to estimate the value of the deformation parameter:
\begin{enumerate}
\item\label{methods1} We can define another parameter $\Delta C$, which is called the deviation from circularity, as $\Delta C=A/(2\langle R \rangle)$.
The EHT collaboration gives the deviation from circularity in the first image of M87* is $\Delta C\lesssim10\%$\cite{EHT_Fi, kerr-hypothesis3}. 
We can get the shadows with different $\delta$ and spin $a$ in the $\delta$-Kerr space-time, then we can calculate the deviation from circularity $\Delta C$ for each shadow to constrain the allowed values of the deformation parameter.
\item\label{methods2} From the areal radius $R_a = \sqrt{A/\pi }$ of the object's shadow we can infer the angular diameter $\theta_d$ on the observer's celestial sky:
\begin{equation}\label{radius}
{{\theta }_{d}}=2\frac{{{R}_{a}}}{d}\ , 
\end{equation}
where $d$ is the distance between M87* and Earth. The ring diameter for the M87* black hole shadow is ${{\theta }_{d}}=42\pm 3\mu as$. Through equation \eqref{radius}, we can get a constrain on the areal radius which can be used to constrain the deformation parameter.
\end{enumerate}

When considering the black hole candidate M87* we need to define the values of its mass and distance from Earth. The mass is usually estimated as ${M}_{\rm BH}\simeq 6.2\times {{10}^{9}}{{M}_{\odot }}$, with $\rm M_\odot$ being one solar mass, while its distance from Earth is usually estimated as $d = 16.8\pm0.8$Mpc \cite{EHT_F,EHT_E} and thus the value of ring diameter is $\theta_d = 42\pm 3\mu as$ \cite{EHT_Fi}.
In addition we need an estimate of the value of the spin and the inclination angle of the disk with respect to Earth. The inclination angle is taken to be $\theta_0 = {17}^{\circ} \pm {2}^{\circ}$ according to \cite{Tamburini_20}.
Then assuming the Kerr geometry the value of the spin has been evaluated as $a=0.90\pm 0.05$ \cite{Tamburini_20}. Of course a non vanishing deformation parameter will affect the measured value of the spin for a given image of the shadow. However, we may expect the departure from the Kerr geometry to be small and therefore the value of the spin to be close to the estimate given in \cite{Tamburini_20}. 
We then wish to investigate how a non vanishing $\delta$ may affect the value of $a$ while keeping $\theta_d$ and $\theta_0$ within the observed ranges. 

\begin{figure}[ttt]
\centering
\includegraphics[width=9.5cm]{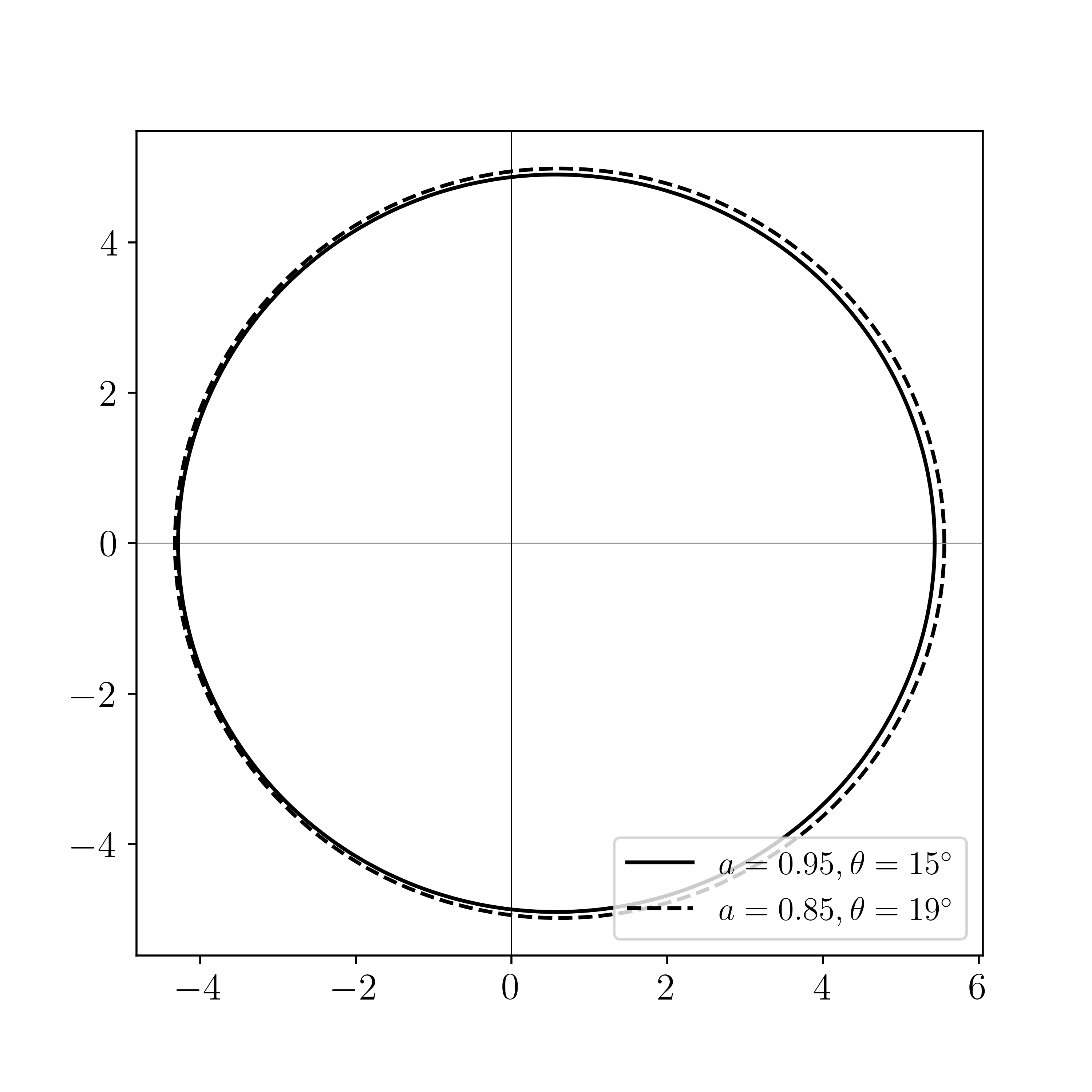}
\caption{The boundary of the shadow in the Kerr metric within experimental bounds from the EHT observations. The two boundaries are obtained for the limiting values of inclination angle and spin of $\theta_0={19}^{\circ}$, $a = 0.85$ and $\theta_0={15}^{\circ}$, $a = 0.95$.}\label{delta_maxmin}
\end{figure}

\begin{figure}[htbp]
\centering
\includegraphics[width=9.5cm]{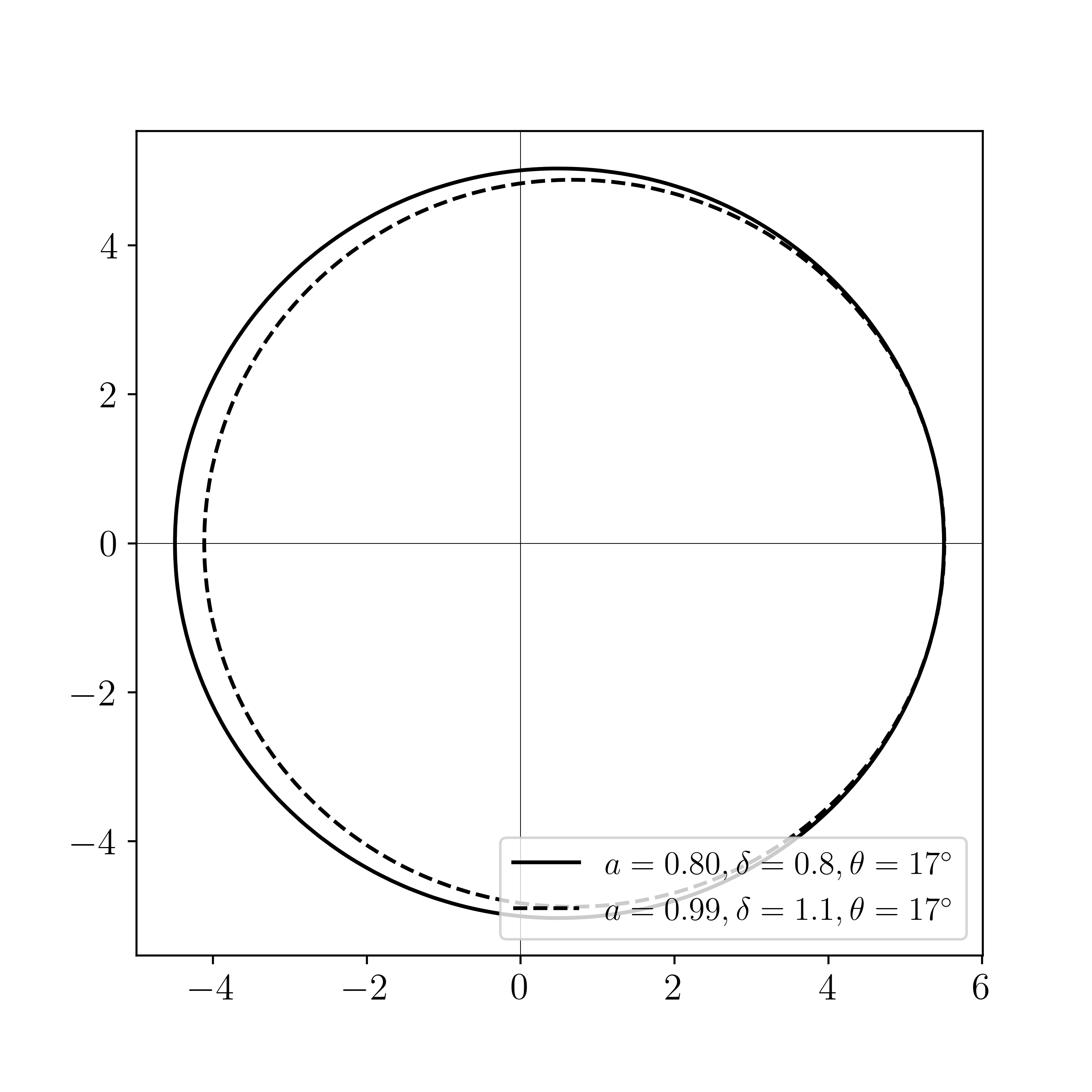}
\caption{The boundary of the shadow in the $\delta$-Kerr metric with fixed inclination angle $\theta_0=17^{\circ}$ and different values for spin $a$ and deformation $\delta$. The two boundaries are obtained for the limiting values of $a=0.80$, $\delta=0.8,$ and $a=0.99$, $\delta=1.1$.}\label{Shadow_new2}.
\end{figure}

Firstly we can notice that for any given value of $\delta$ increasing the value of inclination angle $\theta_0$ (with the spin fixed in the range $a\in (0.85,0.95)$), the area of the shadow increases. Additionally by increasing the value of the spin $a$, with the inclination angle fixed in the range $\theta_0\in ({15}^{\circ},{19}^{\circ})$, the area of the shadow decreases. So we can conclude that, for the Kerr geometry, the minimum radius of the shadow's boundary will be obtained for $\theta_0={19}^{\circ}$, $a = 0.85$, while the maximum value for shadow's boundary will be obtained for $\theta_0={15}^{\circ}$, $a = 0.95$, as can be seen in Fig.~\ref{delta_maxmin}. A similar boundary may be obtained for a fixed inclination angle, by changing the value of the deformation parameter, as can be seen in Fig.~\ref{Shadow_new2}. The shadow's size increases as $\delta$ and $a$ decrease. Therefore within the range of allowed values for the observed angular diameter of the object we may fit different pairs of $\{a, \delta\}$.

\begin{figure}[ttt]
\centering
\includegraphics[width=9.5cm]{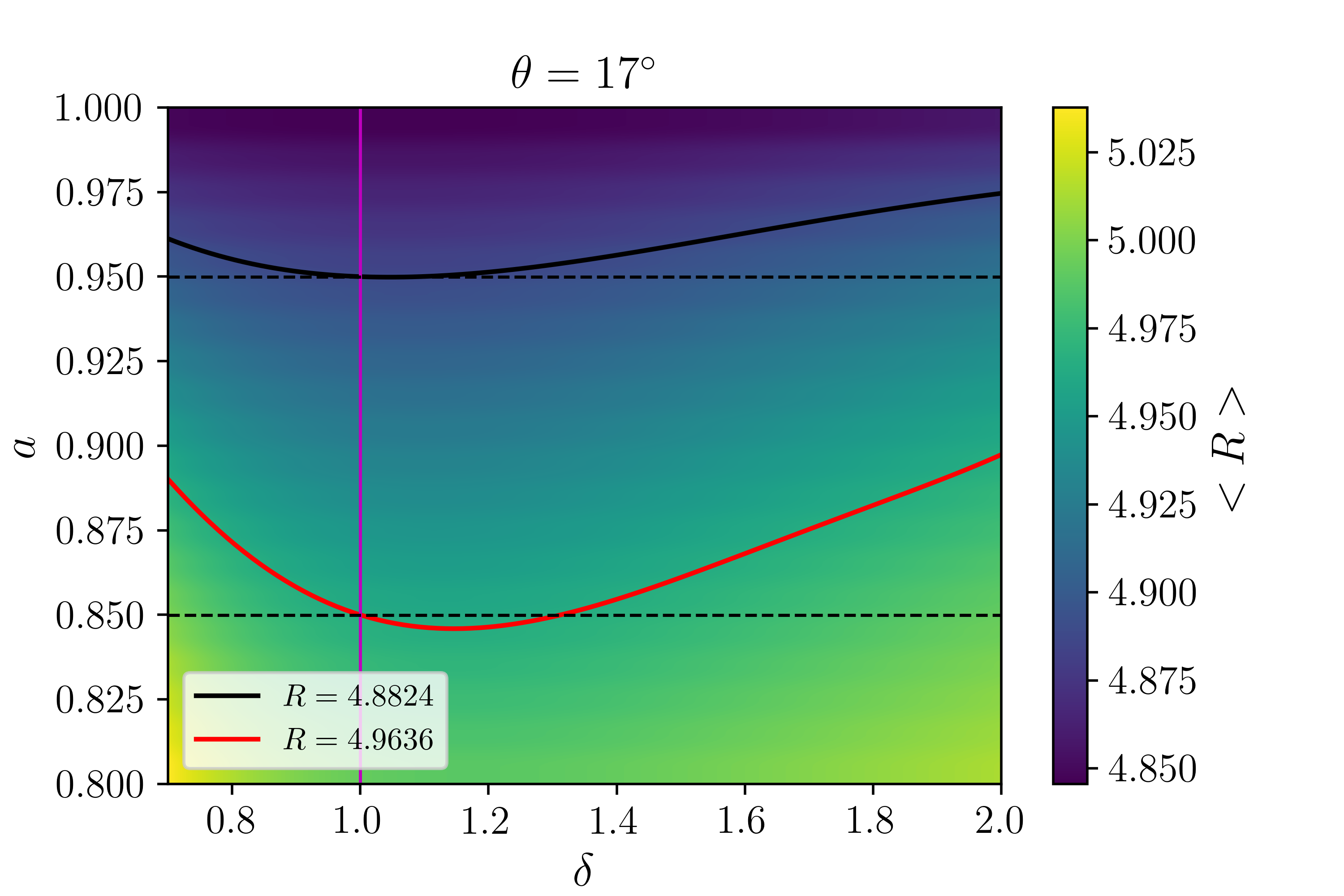}

\caption{The average radius of the shadow $\langle R \rangle$ for different values of spin $a$ and quadrupole parameter $\delta$ with fixed value of observation angle $\theta_0 = 17^{\circ}$. The solid lines correspond to $\langle R \rangle$ for the minimum and maximum values obtained in Fig.~\ref{delta_maxmin}. The purple vertical line corresponds the case of Kerr metric.}\label{Delta_Spin}
\end{figure}

In order to study the effect of the quadrupole parameter on the measurements of the spin of the compact object we considered the constraints on the shadow's boundary $\langle R \rangle$ under the assumption of the Kerr geometry, i.e. Fig.~\ref{delta_maxmin}. To this aim we determined for each value of $\delta$ which range of values of $a$ would give the same constraints on $\langle R \rangle$.
This is shown for three values of the inclination angle $\theta_0$ in the allowed range in Figs~\ref{Delta_Spin}.
The solid black and red lines in the figures represent the shadow's boundary radius values of $\langle R \rangle = 4.8824$ and $\langle R \rangle = 4.9636$, which correspond to the solid and dashed lines in Fig.~\ref{delta_maxmin}. The purple vertical line corresponds the case of the Kerr metric.  
Through these figures one may come to the following conclusions:
\begin{itemize}
    \item[(i)] A slightly oblate ($\delta>1$) black hole mimicker would produce the same shadow with a slightly lower angular momentum with respect to the Kerr case.
	\item[(ii)] For angular momentum close to the observed range $a\simeq 0.9$ the influence of the quadrupole parameter is small, as can be seen from equations \eqref{m0} and \eqref{j1}. Even when the value of $\delta$ becomes large, i.e. $\delta\simeq 2$, the difference of shadow's radius $\langle R \rangle$ between Kerr and $\delta$-Kerr remains small.
	\item[(iii)] The shadow's radius for a fixed $\delta$ increases as the spin $a$ decreases, consistently with what happens in the Kerr case. However, the influence of the quadrupole parameter $\delta>1$ is not monotonic and the shadow's radius, for a given spin $a$, will reach its maximum value for $\delta<1.2$. 
\end{itemize}

\section{Thin accretion disk in $\delta$-Kerr metric}\label{disk}

We now turn the attention to the simulation of a geometrically thin infinite accretion disk in the $\delta$-Kerr metric following the framework developed for black holes in \cite{Infinite_Disk}. To this purpose we used the open source ray-tracing code Gyoto \cite{Gyoto}, which through the ray tracing of photons emitted by matter accreting onto a central object in a given geometry allows to simulate the disk's image.

The original Gyoto code for geometrically thin infinite accretion disks was developed in order to produce images of the shadow of a Kerr black hole. Therefore we had to modify the code in order to apply it to the $\delta$-Kerr metric. The two major changes involve of course the metric functions and the equations of motion. 
Changing the metric functions in the code is straightforward, while the equations of motion required a little more care.
From the Lagrangian for test particles 
\begin{equation}
    \mathcal{L}= \frac{1}{2} g_{\alpha \beta} \dot{q}^{\alpha} \dot{q}^{\beta},
\end{equation}
we get the conjugate momenta as
\begin{equation}
{{p}^{\alpha }} \equiv {{\dot{q}}^{\alpha }}, \quad
p_{\alpha}  \equiv \frac{\partial \mathcal{L}}{\partial \dot{q}^{\alpha}} 
=g_{\alpha \beta} \dot{q}^{\beta}=g_{\alpha \beta} p^{\beta} .
\end{equation}
Notice that the line-element is stationary and exhibits invariance under rotations about the symmetry axis. Therefore, as we mentioned earlier, there are two conserved quantities related to those symmetries, namely the energy per unit mass of the test particle, $E = -p_{t}$, and the component of the angular momentum along the symmetry axis, $ L = -p_{\phi}$.
We can then define the Hamiltonian from the usual Legendre transformation
\begin{equation}
H=p_{\alpha} \dot{q}^{\alpha}-\mathcal{L}=\frac{1}{2} g^{\alpha \beta} p_{\alpha} p_{\beta},
\end{equation}
and write the equations of motion in the form of Hamilton's equations:
\begin{equation}
\dot{q}_{\alpha}=\frac{\partial H}{\partial p_{\alpha}} , \quad \dot{p}_{\alpha}=-\frac{\partial H}{\partial q_{\alpha}} \; .
\end{equation}
Of the above set of equations two (for $t$ and $\phi$) retrieve the conserved quantities while two (for $r$ and $\theta$) are the relevant ones. They are

\begin{equation}\label{Hamilto}
\begin{aligned}
{{{\dot{p}}}_{r}}
&=-\frac{1}{2}({g^{tt}})'{{E}^{2}}-\frac{1}{2}({g^{rr}})'{{p}_{r}}^{2}-\frac{1}{2}({g^{\theta\theta}})'{{p}_{\theta }}^{2}+\\
&-\frac{1}{2}({g^{\phi \phi}})'{{L}^{2}}-({g^{t \phi}})'EL, \\ 
{{{\dot{p}}}_{\theta }}
&=-\frac{1}{2}({g^{tt}})_{,\theta}{{E}^{2}}-\frac{1}{2}({g^{rr}})_{,\theta}{{p}_{r}}^{2}-\frac{1}{2}({g^{\theta\theta}})_{,\theta}{{p}_{\theta }}^{2}+\\
&-\frac{1}{2}({g^{\phi \phi}})_{,\theta}{{L}^{2}}-({g^{t \phi}})_{,\theta}EL, \\
\end{aligned}
\end{equation}
where the superscript $x'$ denotes differentiation with respect to $r$ and now we use the notation $\dot{x}$ to denote differentiation with respect to $t$. Also the subscript $x_{,\theta}$ denotes differentiation with respect to $\theta$.

An important element in constructing the thin disk's image is the profile of the emitted flux of radiation, such as it was described for example in \cite{emitt}, that needs to be implemented in the code.
For a massive test particle orbiting the $\delta$-Kerr metric on a circular orbit in the equatorial plane $\theta=\pi/2$ the equation of motion \eqref{Hamilto} vanishes and the particle's effective potential can be written as:
\begin{equation}
V_{\mathrm{eff}}(r)=-1+\frac{E^{2} g_{\phi \phi}+2 E L g_{t \phi}+L^{2} g_{t t}}{g_{t \phi}^{2}-g_{t t} g_{\phi \phi}}.
\end{equation}
Where the conserved quantities $E$ and $L$ 
can be expressed as:
\begin{equation}
E=-\frac{g_{t t}+\Omega g_{t \phi}}{\sqrt{-g_{t t}-2 \Omega g_{t \phi}-\Omega^{2} g_{\phi \phi}}},
\end{equation}
\begin{equation}
L=\frac{\Omega g_{\phi \phi}+g_{t \phi}}{\sqrt{-g_{t t}-2 \Omega g_{t \phi}-\Omega^{2} g_{\phi \phi}}},
\end{equation}
with the angular velocity $\Omega$ given by
\begin{equation}
\Omega=\frac{d \phi}{d t}=\frac{-g_{t \phi, r} \pm \sqrt{-\left(g_{t \phi, r}^{2}+g_{\phi \phi, r} g_{t t, r}\right)}}{g_{\phi \phi, r}}.
\end{equation}
The inner edge of the thin accretion disk in the $\delta$-Kerr space-time is given by the innermost stable circular orbit (ISCO). The position of the ISCO is obtained from the conditions:
\begin{equation}
V_{\mathrm{eff}}(r)= V_{\mathrm{eff}}^{\prime}(r)= V_{\mathrm{eff}}^{\prime \prime}(r)=0,
\end{equation}

\begin{figure}[htbp]
\centering
\includegraphics[width=8.5cm]{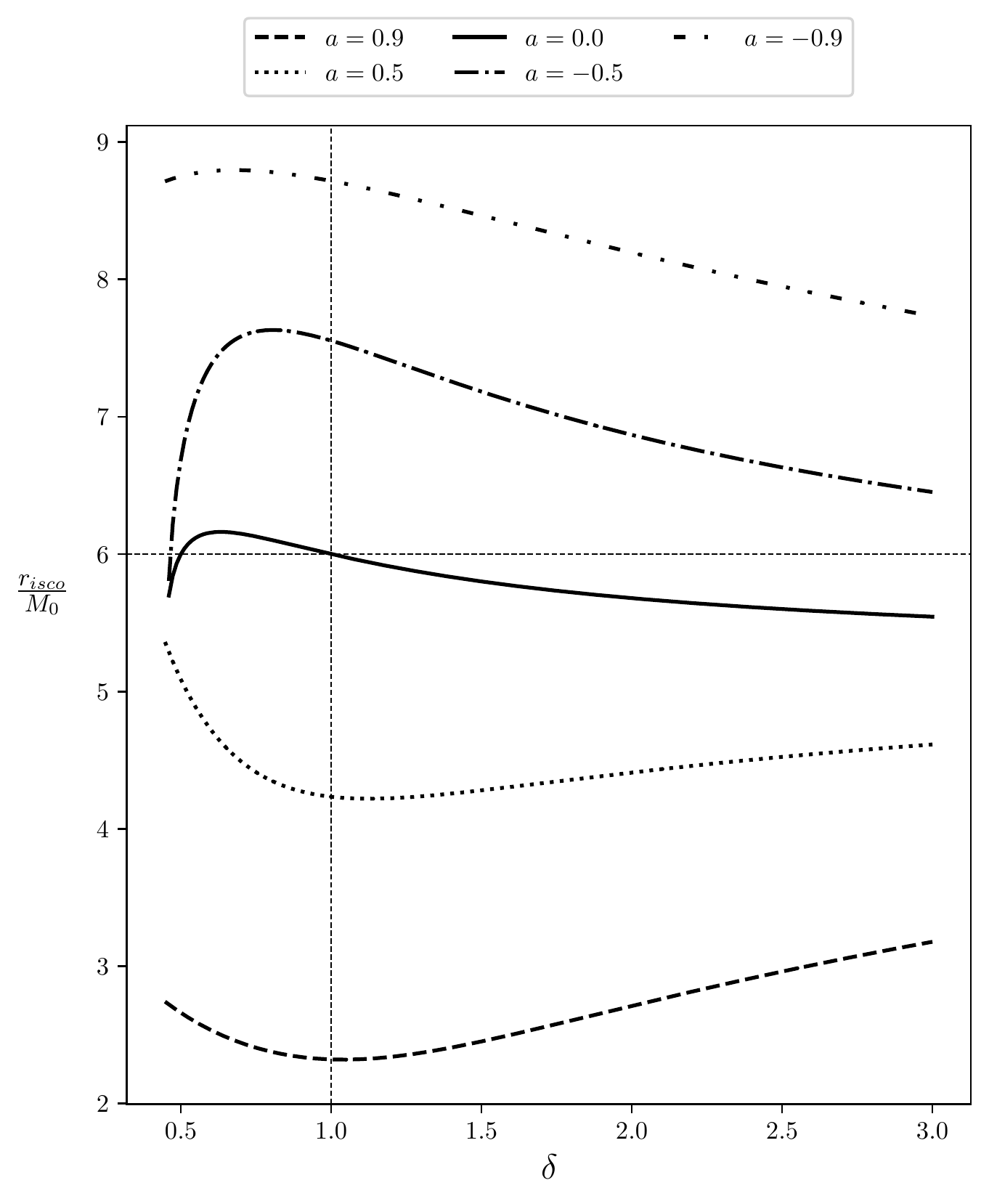}
\caption{ The radius of the innermost stable circular orbit (ISCO) in units of $M_0$ as a function of the deformation parameter $\delta$ for different values of the spin parameter $a$. The solid line corresponds to the Zipoy-Voorhees metric. The vertical line for $\delta=1$ corresponds to the Kerr case.}\label{ISCO}
\end{figure}

which determine the energy $E$, angular momentum $L$ and location $r_{\rm ISCO}$ of the particle on the innermost stable circular orbit. In Fig.~\ref{ISCO} we show the dependence of the ISCO on the deformation parameter $\delta$ for various values of $a$ in units of the object's mass $M_0$.
Assuming that the accretion disk emits electromagnetic radiation, the total radiative flux can be defined as
\begin{equation}
F(r)=-\frac{\dot{M}_{0} c^{2}}{4 \pi M_0^{2}} \frac{\Omega_{, r}}{\sqrt{-g}(E-\Omega L)^{2}} \int_{r_{I S C O}}^{r}(E-\Omega L) L_{, r} d r,
\end{equation}
where $\dot{M}_0$ is the mass accretion rate, which can be approximated as constant and $g$ is the determinant of the three dimensional sub-space metric with $\theta=\pi/2$ of the $\delta$-Kerr metric.

\begin{figure*}
\centering
\includegraphics[width=18cm]{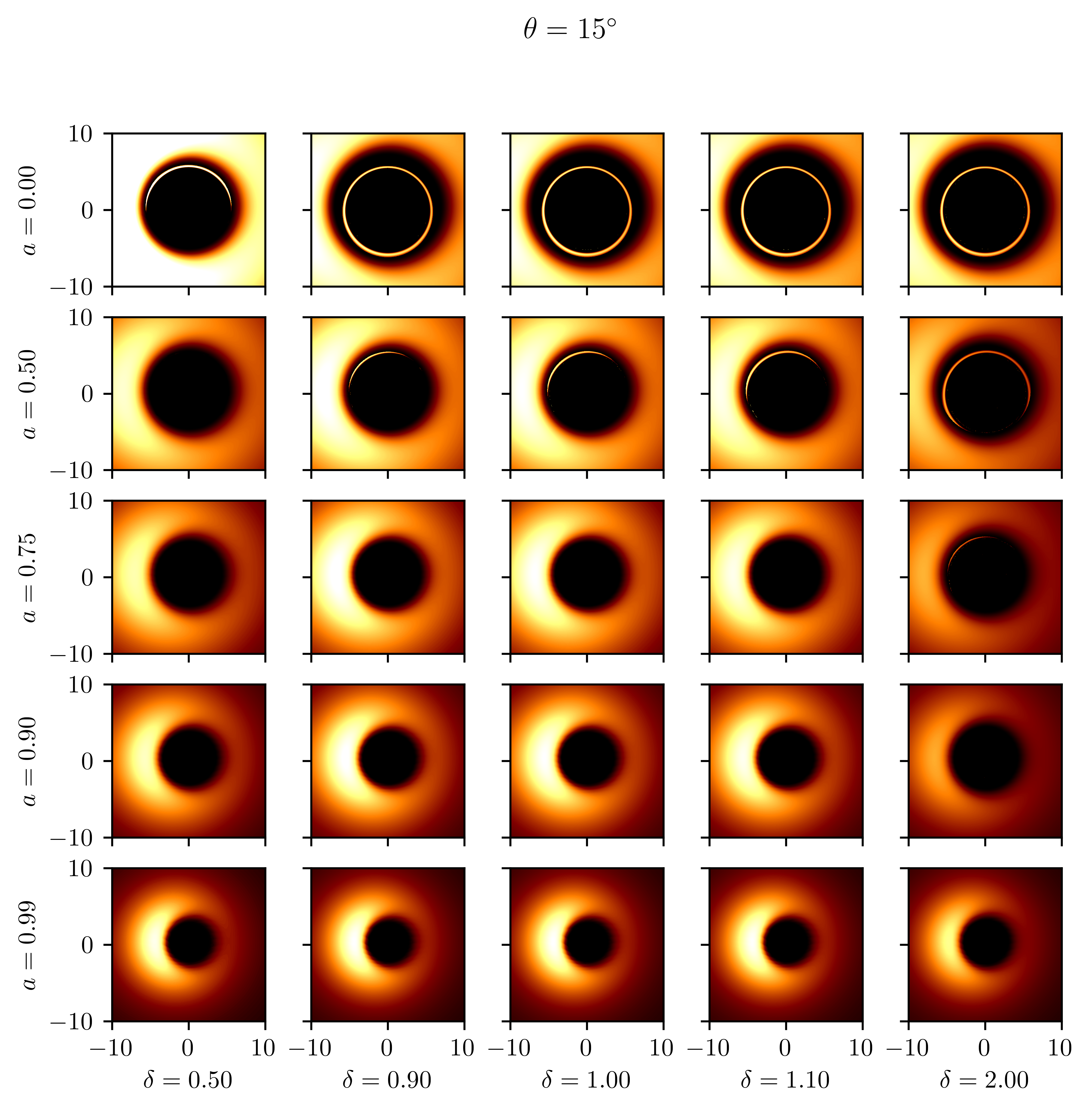}
\caption{The thin accretion disk in the $\delta$-Kerr space-time with the observation angle $\theta_0 = 15^{\circ}$, for the different values of $a$ and $\delta$.
Image sizes are normalized with $M_0$ while the energy flux of each rows (fixed $a$) is normalized with the maximum value in the Kerr case ($\delta = 1$). The top row corresponds to the Zipoy-Voorhees metric, while the middle column corresponds to the Kerr case.
}\label{acc_15}
\end{figure*}

\begin{figure*}
\centering
\includegraphics[width=18cm]{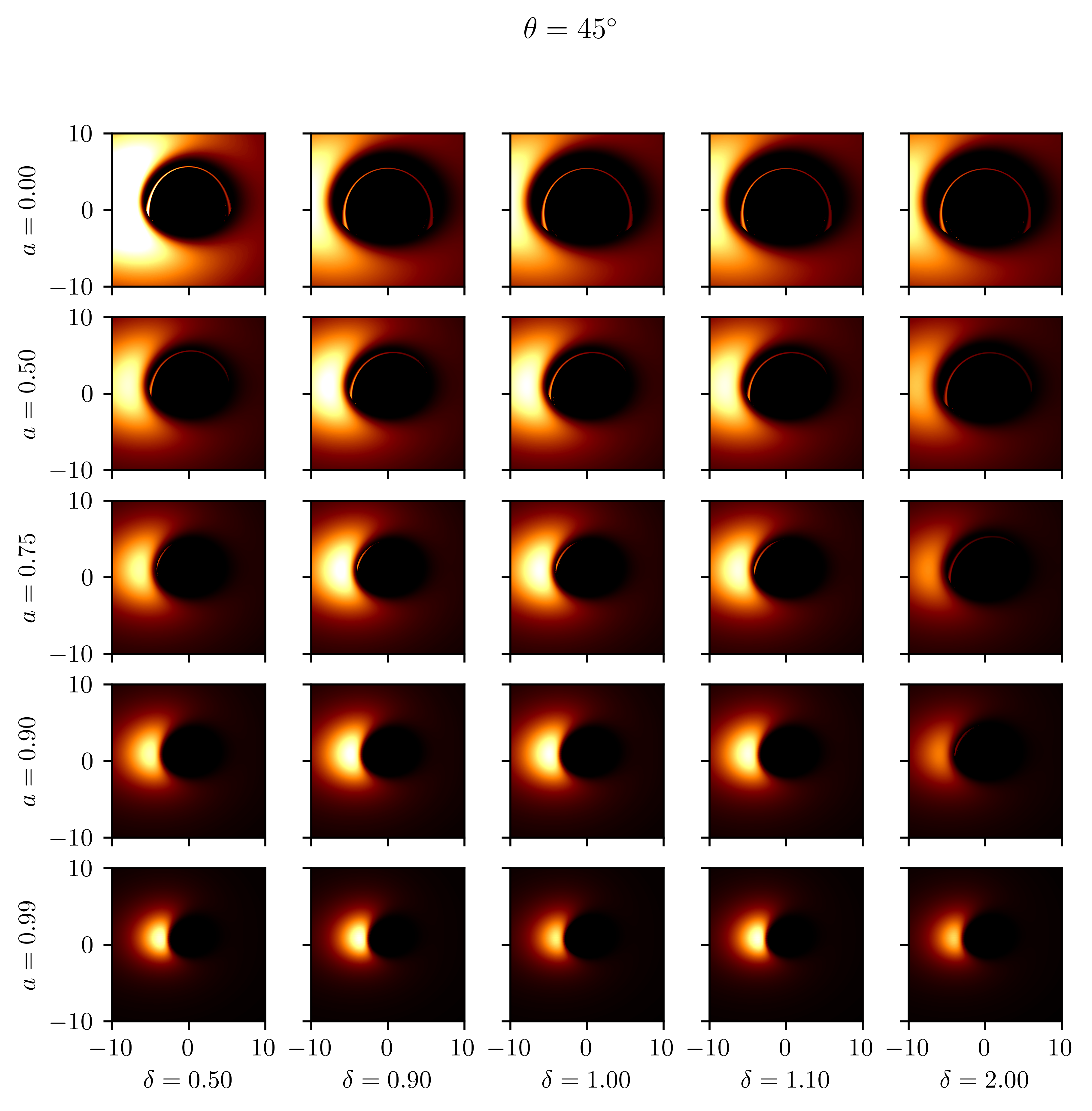}
\caption{The thin accretion disk in the $\delta$-Kerr space-time with the observation angle $\theta_0 = 15^{\circ}$, for the different values of $a$ and $\delta$.
Image sizes are normalized with $M_0$ while the energy flux of each rows (fixed $a$) is normalized with the maximum value in the Kerr case ($\delta = 1$). The top row corresponds to the Zipoy-Voorhees metric, while the middle column corresponds to the Kerr case.}\label{acc_45}
\end{figure*}

\begin{figure*}
\centering
\includegraphics[width=18cm]{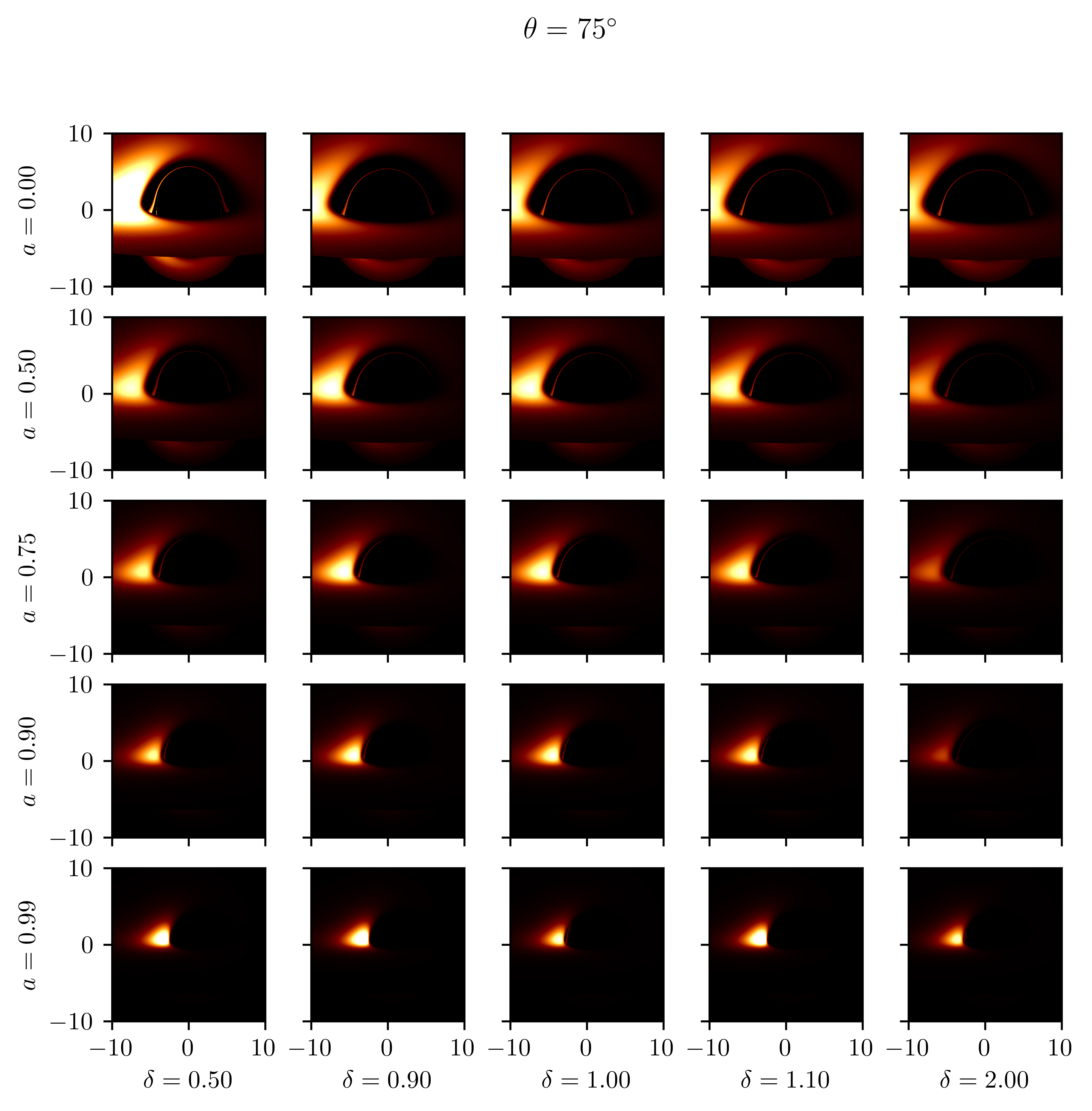}
\caption{The thin accretion disk in the $\delta$-Kerr space-time with the observation angle $\theta_0 = 15^{\circ}$, for the different values of $a$ and $\delta$.
Image sizes are normalized with $M_0$ while the energy flux of each rows (fixed $a$) is normalized with the maximum value in the Kerr case ($\delta = 1$). The top row corresponds to the Zipoy-Voorhees metric, while the middle column corresponds to the Kerr case.}\label{acc_70}
\end{figure*}

\begin{figure*}
\centering
\includegraphics[width=18cm]{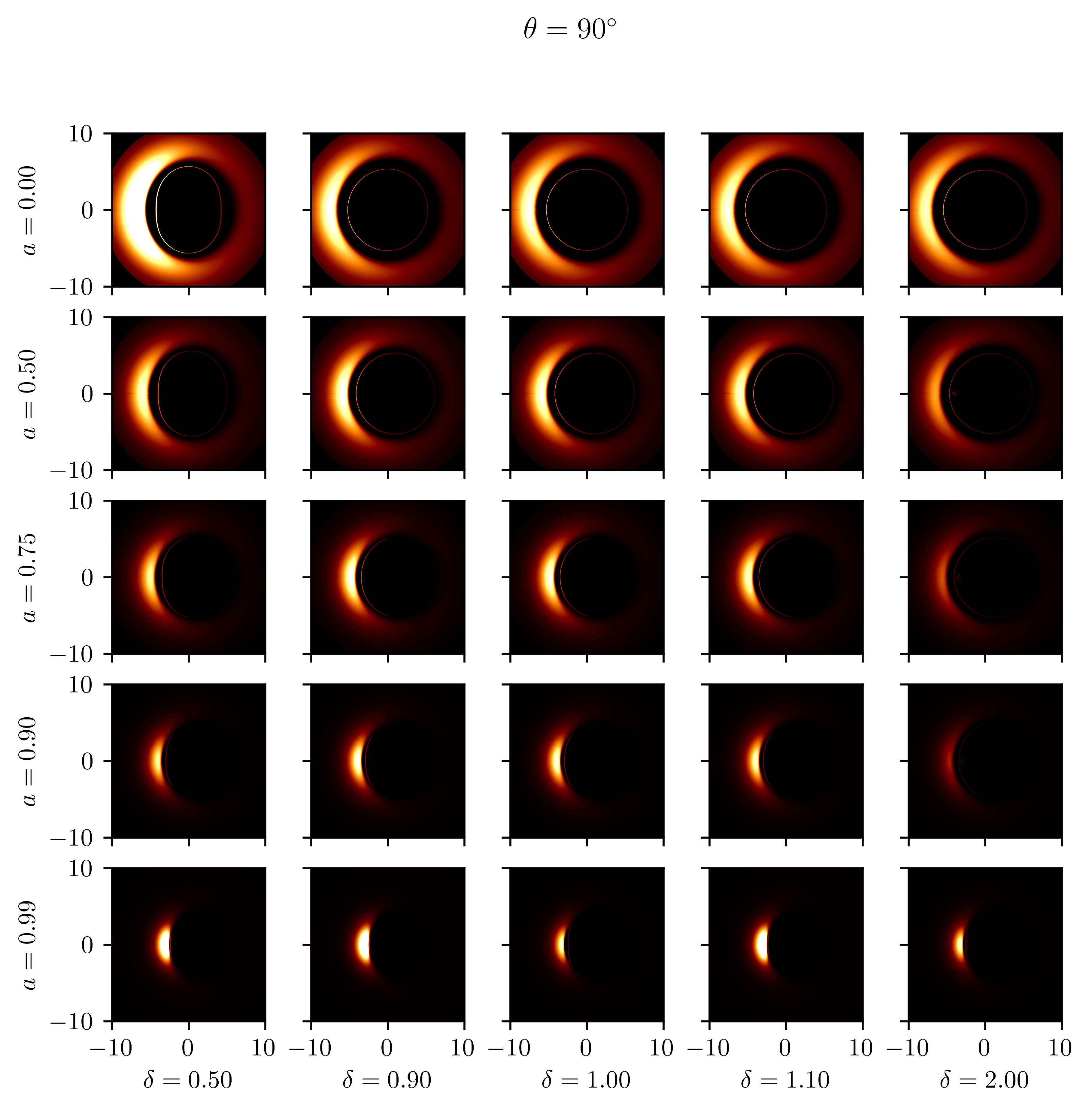}
\caption{The thin accretion disk in the $\delta$-Kerr space-time with the observation angle $\theta_0 = 15^{\circ}$, for the different values of $a$ and $\delta$.
Image sizes are normalized with $M_0$ while the energy flux of each rows (fixed $a$) is normalized with the maximum value in the Kerr case ($\delta = 1$). The top row corresponds to the Zipoy-Voorhees metric, while the middle column corresponds to the Kerr case.}\label{acc_90}
\end{figure*}

In Figs~\ref{acc_15}, \ref{acc_45}, \ref{acc_70}, \ref{acc_90} we show the image of the thin accretion disk in the $\delta$-Kerr space-time for different values of $a$ and $\delta$ at inclination angles $\theta_0$ of $15^\circ$,  $45^\circ$, $70^\circ$ and $90^\circ$. 
The middle column of the figures shows the thin accretion disk in the Kerr case, i.e. $\delta=1$,  and it is clear that the most significant departures are obtained for prolate sources ($\delta <1$) with small values of $a$.
Therefore for the range of values constrained by the image of M87* it may not be possible to distinguish whether the image is due to a Kerr black hole or to a $\delta$-Kerr metric with slightly different values of the parameters, see Fig. \ref{acc_15} with $a=0.9$, for example. In fact, as suggested by the analysis of Figs~\ref{Delta_Spin}, for large angular momentum even larger values of $\delta$ may produce the same image of the disk. It is then clear that the fact that M87* may be a black hole mimicker described by the $\delta$-Kerr metric can not be excluded based solely on the properties of the shadow as inferred from a unique set of observations. Therefore independent measurements of the deformation and spin of the object may be necessary in order to break the degeneracy and determine the nature of the geometry outside the compact object.

\section{Conclusion\label{Summary}}

We investigated the possibility to test the Kerr hypothesis via the image of the shadow of supermassive black holes. To this aim we considered a black hole mimicker whose exterior geometry is given by the $\delta$-Kerr metric, an exact solution of the field equations in vacuum which is continuously linked to the Kerr metric via the value of one parameter.
The $\delta$-Kerr space-time is a generalization of the Kerr space-time which belongs to the class of Ricci-flat exact solutions of gravitational field equations. It is obtained from the nonlinear superposition of the Zipoy-Voorhees metric and the Kerr metric and it can be thought of as describing a deformed Kerr black hole mimicker with the deformation parameter given by $\delta = 1 + q$, where $q \ne 0$ is related to the non relativistic mass quadrupole moment of the source. The presence of the additional parameter turns the Kerr horizon into a curvature singularity thus suggesting the interpretation that the line element describes the exterior of an exotic compact object with boundary slightly larger that the infinitely redshifted surface.

We studied the optical properties of $\delta$-Kerr space-time with a ray-tracing code for photons and simulated the apparent shape of the shadow of the compact object with the aid of the Gyoto code \cite{Gyoto}.
We showed that the shadow of the black hole candidate M87* obtained by the EHT collaboration \cite{EHT_E} could be produced also by an accretion disk in the $\delta$-Kerr space-time with non vanishing deformation $q$ and values for the angular momentum and inclination angle within the measured range. Therefore a separate set of measurements of those quantities with an independent method is necessary in order to break the degeneracy and determine the nature of the geometry.

We conclude that the image of the shadow alone, however accurate, is not enough to exclude this class of black hole mimickers since within the error bars of the measurements it will always be possible to find sets of $\{a,\delta\}$ that produces the same image. However the degeneracy may be broken by an independent set of measurements of the same parameters, for example by measuring the orbits of test particles, such as stars orbiting the supermassive black hole candidate, in the vicinity of the object's ISCO. As of now the shadow of supermassive BH candidates has been imaged for M87* and SgrA*. In the case of SgrA* we already know the orbits of some nearby stars and its relative vicinity to Earth may one day lead to the discovery of neutron stars orbiting close enough to the ISCO to allow us to break the degeneracy in the measurement of the deformation parameter and determine if the geometry is indeed given by the Kerr metric.

Another way to obtain an independent constraint of the deformation parameter of the $\delta$-Kerr metric may be through the observation of gravitational wave signals. This may be impossible at present for M87* and SgrA*, as we do not expect to be able to detect gravitational waves from the inspiral of stars and neutron stars onto the central object. However it may be possible to numerically study the merger of a binary stellar mass system described by two $\delta$-Kerr compact objects and compare the results with the available data from LIGO and Virgo. Additionally constraints on the allowed values of the deformation parameter may come from the study of the emission spectra produced by the disks surrounding supermassive black hole candidates in the universe.

In the last few years the number and variety of available observations of the near horizon region of black hole candidates has grown significantly. To the point that soon we will have sufficient constraints on the exterior geometry of such compact objects to test the validity of the Kerr hypothesis. Until that time it is important to study the properties of black hole mimickers such as the $\delta$-Kerr metric.

\section*{acknowledgements}
This research is supported by The National Key R\&D Program
of China (Grant No. 2021YFC2203002), Grants F-FA-2021-432 and MRB-2021-527 of the Uzbekistan Ministry for Innovative Development and
by the Abdus Salam International Centre for Theoretical Physics under the Grant No. OEA-NT-01.. A. A. is supported by the PIFI
fund of Chinese Academy of Sciences.
DM acknowledges support by Nazarbayev University Faculty Development Competitive Research Grant No. 11022021FD2926. W. H. is supported by CAS Project for Young Scientists in Basic Research YSBR-006, NSFC (National Natural Science Foundation of China) No. 11773059 and No. 12173071.

\bibliographystyle{apsrev4-1} 
%


\end{document}